\documentclass[prl, twocolumn,superscriptaddress,nofootinbib]{revtex4}
\usepackage{amsmath,amssymb,bm}
\usepackage{graphicx}
\usepackage{epstopdf}
\usepackage{latexsym}
\usepackage{subfigure}
\usepackage{color}
\usepackage{natbib}
\usepackage{braket}
\usepackage{hyperref}
\hypersetup{
  colorlinks,
  citecolor=magenta,
  linkcolor=blue,
  urlcolor=blue}
\newcommand{\nlio}{(Na$_{1-x}$Li$_{x}$)$_2$IrO$_3$}
\newcommand{\nkio}{(Na$_{1-x}$K$_{x}$)$_2$IrO$_3$}
\newcommand{\naio}{(Na$_{1-x}$A$_{x}$)$_2$IrO$_3$}
\newcommand{\nio}{Na$_2$IrO$_3$}
\newcommand{\kio}{K$_2$IrO$_3$}
\newcommand{\lio}{Li$_2$IrO$_3$}
\newcommand{\jeff}{J_{\rm eff}}

\begin{document}

\title{Evolution of Magnetism in Single-Crystal Honeycomb Iridates}

\author{G.~Cao}
\author{T.~F.~Qi}
\author{L.~Li}
\author{J.~Terzic}
\affiliation{Department of Physics \& Astronomy, University of Kentucky, Lexington, KY 40506-0055}
\author{V.~S.~Cao}
\affiliation{Paul Laurence Dunbar High School, Lexington, KY 40513}
\affiliation{Department of Physics \& Astronomy, University of Kentucky, Lexington, KY 40506-0055}
\author{S.~J.~Yuan}
\affiliation{Department of Physics \& Astronomy, University of Kentucky, Lexington, KY 40506-0055}
\affiliation{Department of Physics, Shanghai University, Shanghai, China}
\author{M.~Tovar}
\author{G.~Murthy}
\author{R.~K.~Kaul}
\affiliation{Department of Physics \& Astronomy, University of Kentucky, Lexington, KY 40506-0055}

\begin{abstract}
We report the successful synthesis of single-crystals of the layered iridate, \nlio, $0\leq x \leq 0.9$, and a thorough study of its  structural, magnetic, thermal and transport properties. The new compound allows a controlled interpolation between \nio~and \lio, while maintaing the novel quantum magnetism of the honeycomb Ir$^{4+}$ planes. The measured phase diagram demonstrates a suppression of the N\'eel temperature, $T_N$, at an intermediate $x$ indicating that the magnetic order in \nio~and \lio~are distinct. X-ray data shows that for $x\approx 0.7$ when $T_N$ is suppressed the most, the honeycomb structure is least distorted, suggesting at this intermediate doping that the material is closest to the spin liquid that has been sought after in \nio~and \lio.  By analyzing our magnetic data with a single-ion theoretical model we also show that the trigonal splitting, on the Ir$^{4+}$ ions changes sign from \nio~to~\lio.
\end{abstract}
\pacs{75.10.Jm, 75.25.Dk, 75.40.Cx}
\date{\today}
\maketitle

{\em Introduction: --} The iridates have recently been recognized as a unique arena for the study of new phases of matter that arise from simultaneously strong electron-electron and spin-orbit interactions. Thus far, the most novel manifestation of this interplay in this family of materials is the  $t^5_{2g}$, $\jeff = 1/2$ Mott insulating state, originally experimentally observed in the layered perovskite, Sr$_2$IrO$_4$~\cite{kim2008:jhalf,moon2008:rp,kim2009:science}.  
The iridates have inspired a large body of theoretical and experimental work~\cite{krempa2014:annurev}, since the $\jeff$ levels have mixed spin and orbital character, which may result in a host of exciting quantum ground states~\cite{wan2011:weyl}.

The interest in this field received a major boost when a theoretical analysis~\cite{jackeli2009:kit} showed that the oxygen mediated superexchange processes between the Ir$^{4+}$ moments in the honeycomb iridates \nio~and \lio~result in the celebrated ``Kitaev'' model (KM) for the $\jeff=1/2$ degrees of freedom, $H_K = K\sum_{\langle ij\rangle} \sigma^\gamma_i\sigma^\gamma_j$, where $\gamma=x,y,z$ denotes a different Pauli matrix for each direction of bond on the honeycomb lattice and $\vec{\sigma_i}$ acts on the $\jeff=1/2$ states on site $i$. The KM can be solved exactly and its ground state is an exotic magnetically disordered quantum ``spin liquid''~\cite{kitaev2006:hc}. However, it is experimentally established that both honeycomb Iridate compounds order magnetically: \nio~orders at $T_N=18$ K~\cite{singh2010:na}, and \lio~orders at $T_N=15$ K~\cite{kobayashi2003:li,singh2012:kitH}. 
There are many theoretical proposals for the interactions supplementary to the Kitaev model that would cause magnetic ordering including, additional exchange processes~\cite{chaloupka2010:kitH, choi2012:nsc_na, kimchi2011:j2j3,singh2012:kitH,price2012:kitH,chaloupka2013:zigzag,kim2013:li_bs}, strong trigonal fields~\cite{bhattacharjee2012:trig}, or weak coupling instabilities~\cite{mazin2012:na_bs}; currently there is no consensus on which of these is correct.

On the experimental side, there are now fairly thorough studies of \nio~ using both momentum resolved resonant inelastic X-ray and neutron scattering techniques that establish the pattern of magnetic ordering to be of an unusual zigzag type~\cite{liu2011:xray_na,ye2012:nsc_na,choi2012:nsc_na,gretarsson2013:hc_xray}. This has been possible in part due to the availability of large single crystals of~\nio. Because of various difficulties in chemical synthesis, no such single crystals are available for \lio~and the detailed magnetic ordering pattern of this compound is still unknown. It is noted that an early study on polycrystal \lio~exhibited no magnetic order above 5 K~\cite{felner2002:li}, but more recent measurements show a magnetic transition at  $T_N=15$ K~\cite{singh2012:kitH}. The conspicuous absence of single crystals of \lio~is clearly a major roadblock in a complete characterization of this material.

In this work we fill the gap in our understanding by the successful synthesis and study of single crystals of \nlio.\footnote{For comparison only, we also present some new results on the \nkio~ (with $x\leq 0.02$); the pure compound \kio has not been synthesized yet.} 
The central findings of our work are as follows: As $x$ is tuned we find from X-ray data that the lattice parameters evolve monotonically from Na to Li, retaining the basic Mott insulating honeycomb structure of the Ir$^{4+}$ planes for all $x$.  Even so, there is a non-monotonic dramatic change in N\'eel temperature $T_N$ with $x$, in which $T_N$ initially decreases from 18 K at $x=0$ to 1.2 K at $x=0.70$ before it rises to 7 K at $x=0.90$, indicating that the magnetic ground states at $x=0$ and $x=1$ are not related linearly, as had been previously suggested~\cite{singh2012:kitH}. X-ray structure data shows that the Ir$^{4+}$ honeycomb lattice is least distorted at $x\approx 0.7$. Interesting, we find the lowest $T_N$ and highest frustration parameter also at $x\approx 0.7$. 
 In addition, the high-temperature anisotropy in the magnetic susceptibility is simultaneously reversed and enhanced upon Li doping, and as a result, the in-plane magnetic susceptibility $\chi_\parallel(T)$ becomes significantly greater than the  perpendicular-to-plane susceptibility, $\chi_\perp(T)$ or $\chi_\parallel(T) > \chi_\perp(T)$ for $x=0.90$, which sharply contrasts with the the weaker magnetic anisotropy for $x=0$ where $\chi_\parallel(T) < \chi_\perp(T)$. Using a single ion theoretical model, we show that the anisotropy arises from a trigonal crystal field $\Delta$ oriented perpendicular to the honeycomb layers, and the anisotropy change is the result of {\em sign change} in the local trigonal field between the Na and Li compounds. We thus conclude that the magnetic ground states of the pure Na and Li compounds are distinct and are separated by a quantum phase transition that is driven by the evolution of the crystal structure as $x$ is varied.

\begin{figure}[t]
\centerline{\includegraphics[trim = 0mm 0mm 0mm 0mm, clip,width=\columnwidth]{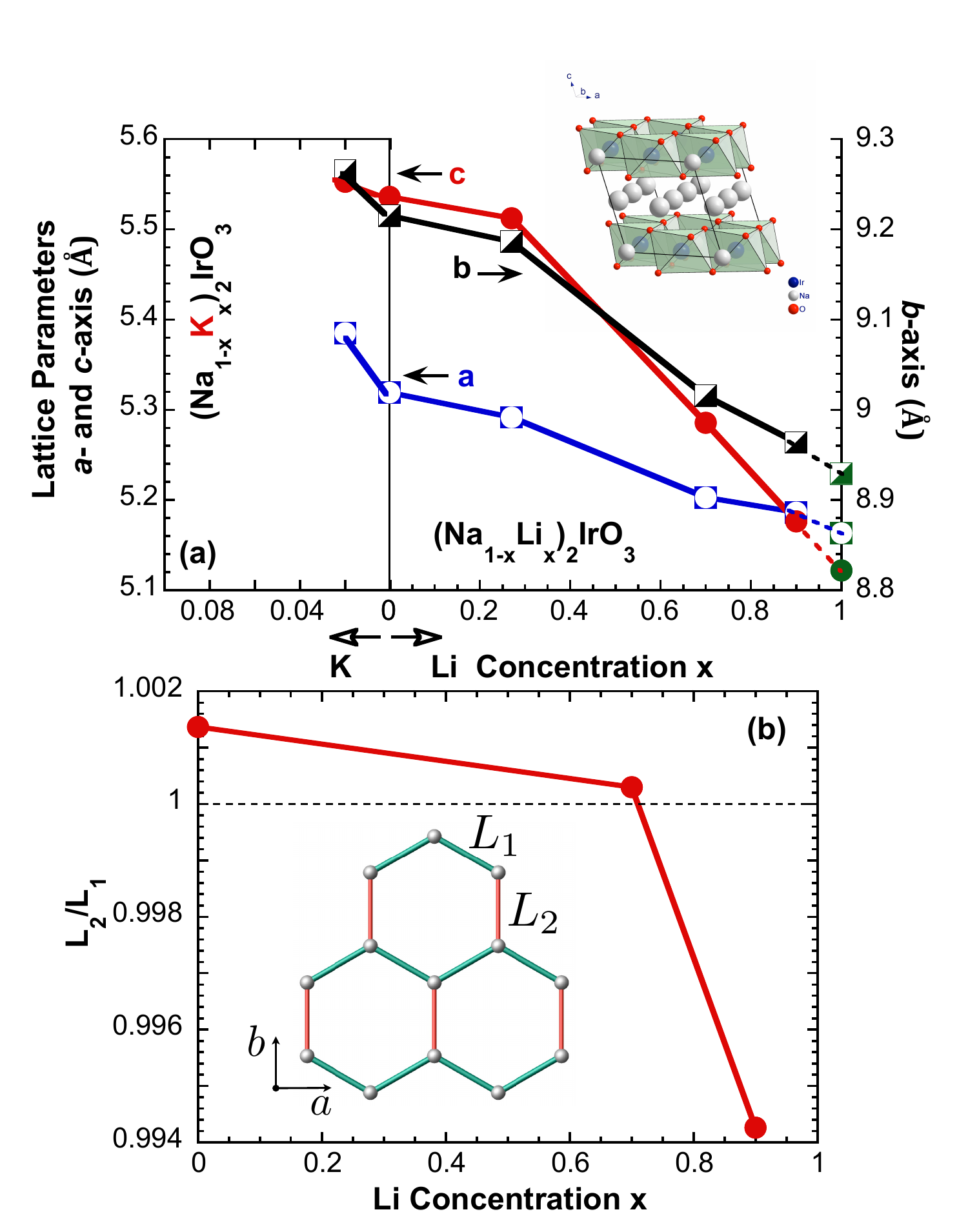}}
\caption{Synthesis and structure of \naio~ (with A= Li or K). (a) The lattice parameters $a$ and $c$ (left scale) and the $b$ (right scale) as a function of $x$. Note that the green data points for \lio~($x=1.0$) are obtained from Ref.~\cite{singh2012:kitH}; {\em Inset:} An illustration of the crystal structure of~\nio. (b) The evolution of distortion of the honeycomb lattice of Ir$^{4+}$ ions from $L_2>L_1$ for $x=0$ to $L_1>L_2$ for $x=1$. At $x\approx0.7$, the system is near perfect honeycomb.}
\label{fig:intro}
\end{figure}

{\em Measurements: --} The methods by which our single crystals are grown and the measurements are carried out are described in the Supplementary Materials. Li doping retains the space group of C2/m that \nio~adopts but induces a systematic decrease in the lattice parameters since the ionic radius of the Li ion is approximately 25\% smaller than that of the Na ion. The lattice parameters are shown in Fig.~\ref{fig:intro}(a).  An important feature of this change is that the lattice parameter $c$ is more severely compressed than the $a$ and $b$. For example, for $x=0.90$, the decrease in the $a$, $b$ and $c$ is 2.5\%, 2.7\% and 6.5\%, respectively. The corresponding angle between the $c$-axis and the basal plane, $\beta$, increases to 109.58$^\circ$ for $x=0.90$ from 108.67$^\circ$ for $x=0$. In Fig.~\ref{fig:intro}(b), we show how the distortion of the $a$-$b$ honeycomb lattice on which the Ir$^{4+}$ moments reside evolves with $x$. The ratio, $L_2/L_1$ (defined in the inset), clearly shows the lattice anisotropy reversal from Na to Li.

\begin{figure}[t]
\centerline{\includegraphics[trim = 5mm 10mm 20mm 10mm, clip,width=\columnwidth]{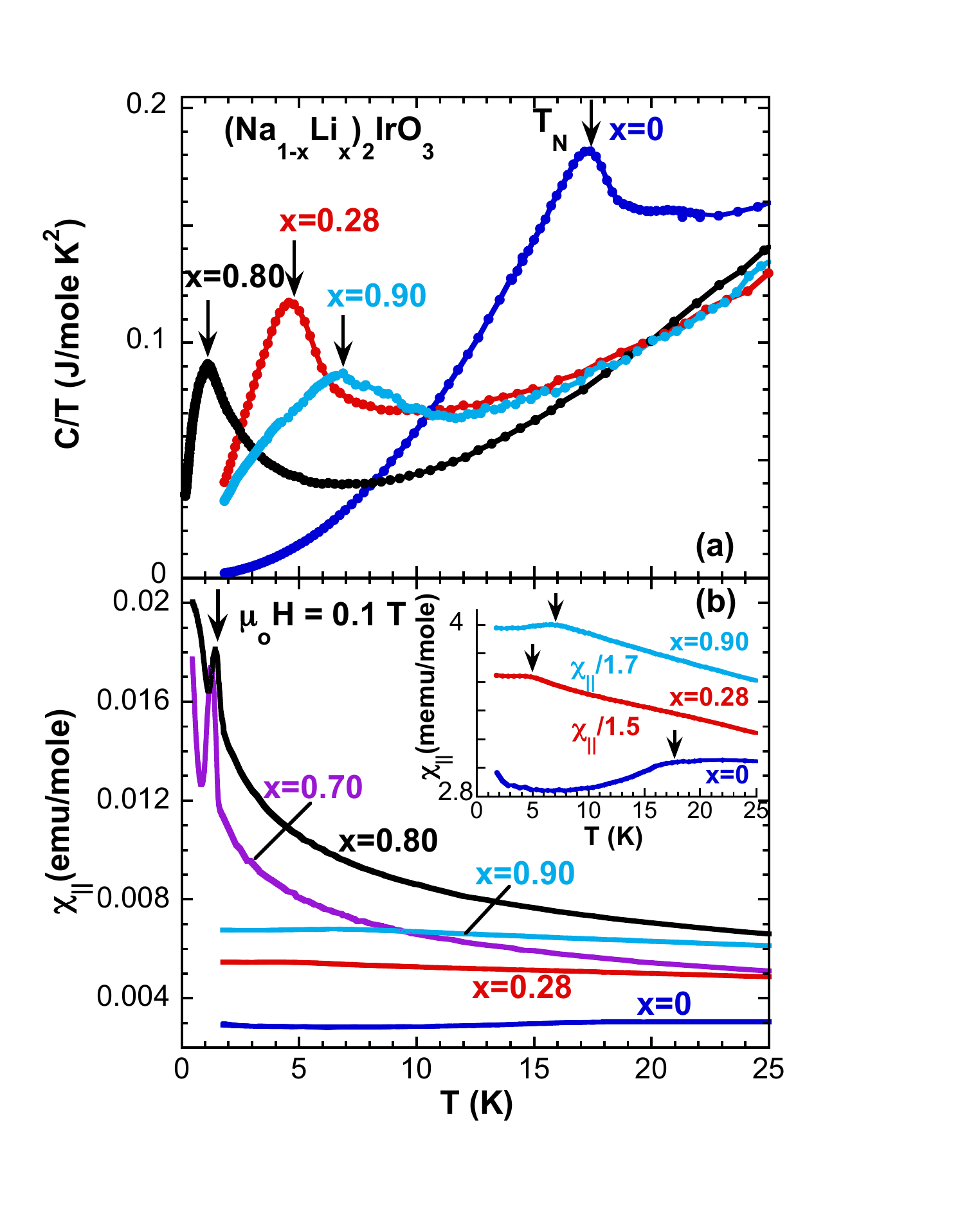}}
\caption{ Study of the N\'eel transition as a function of $x$ in \nlio~in the specific heat and in-plane susceptibility. (a) The specific heat for different $x$. We have plotted $C(T)/T$ for presentation purposes. (b) The in-plane magnetic susceptibility $\chi_\parallel(T)$ at $\mu_0 H = 0.1$ T. {\em Inset:} Zoom-in of the $\chi_\parallel$ data to show kinks at the phase transitions. }
\label{fig:cvchi}
\end{figure}

\begin{figure}[t]
\centerline{\includegraphics[trim = 5mm 10mm 10mm 10mm, clip,width=\columnwidth]{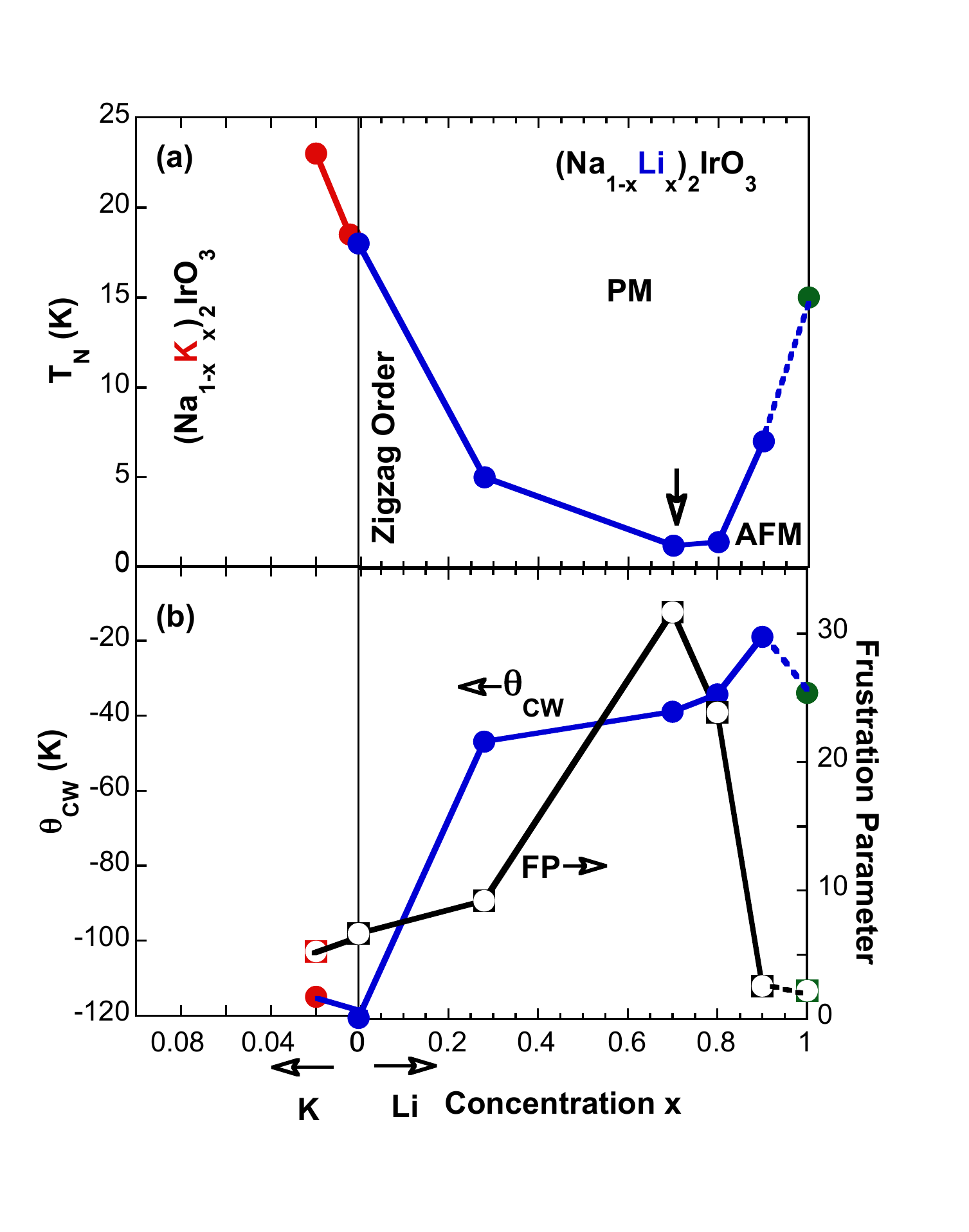}}
\caption{Ordering and interaction scales of \naio~(A- Li or K) as a function of $x$. (a) The N\'eel temperature $T_N$, and (b) the Curie-Weiss temperature $\theta_{\rm CW}$ (left scale) and the frustration parameter (right scale) as a function of $x$. Note that K doping increases TN in sharp contrast to Li doping and that the green data points for \lio~($x=1.0$) are obtained from Ref.~\cite{singh2012:kitH}. }
\label{fig:pd}
\end{figure}

We extract $T_N$ by locating the low temperature peak in the specific heat $C(T)$, shown in Fig.~\ref{fig:cvchi}(a), as well as the in-plane susceptibility, $\chi_\parallel(T)$, shown in Fig.~\ref{fig:cvchi}(b). The pronounced peaks in $C(T)/T$ unambiguously signal a continuous magnetic phase transition for all $x$. As shown in the main panel and inset of Fig.~\ref{fig:cvchi}(b) we find consistent values for $T_N$ extracted from $\chi_\parallel(T)$. Interesting $T_N(x)$ is not a smooth interpolation between the already known $x=0$ and $x=1$ limits.  It is initially suppressed from 18 K for $x=0$ to 5 K for $x=0.28$ and then to 1.2 K for $x=0.70$ before it rises to 1.4 K for $x=0.80$ and finally 7 K for $x=0.90$. We note that the trend of $T_N$ increasing again in our single crystal $x=0.9$ is consistent with previous measurements on polycrystalline samples of \lio~\cite{singh2012:kitH}. A phase diagram that summarizes $T_N$ and the Curie-Weiss scale $\theta_{\rm CW}$ as a function of $x$ is shown in Fig.~\ref{fig:pd}(a,b). A natural conclusion based on the strongly suppressed value of $T_N$ is that it goes to zero for some $x$ and one encounters at least one quantum phase transition in the evolution from \nio~to \lio~at $x\approx 0.7$, implying that the magnetic ground states of \nio~and \lio~are not adiabatically connected. It is interesting to note that at $x\approx 0.7$ when $T_N$ is suppressed the most, the honeycomb plane is closest to ideal ({\em i.e.}, $L_2/L_1\approx 1$ in Fig.~\ref{fig:intro}(b)). Finally, we note that the dominant role of Li doping is to tune the chemical pressure, which in turn causes an evolution of the lattice structure. This can be verified by noting that if we dope in K instead of Li, $T_N$ {\em increases}, as shown in Fig.~\ref{fig:pd} (a). This is consistent with K doping achieving negative chemical pressure because the ionic radius of K is larger than Na.


\begin{figure}[t]
\centerline{\includegraphics[trim = 10mm 20mm 20mm 10mm, clip,width=\columnwidth]{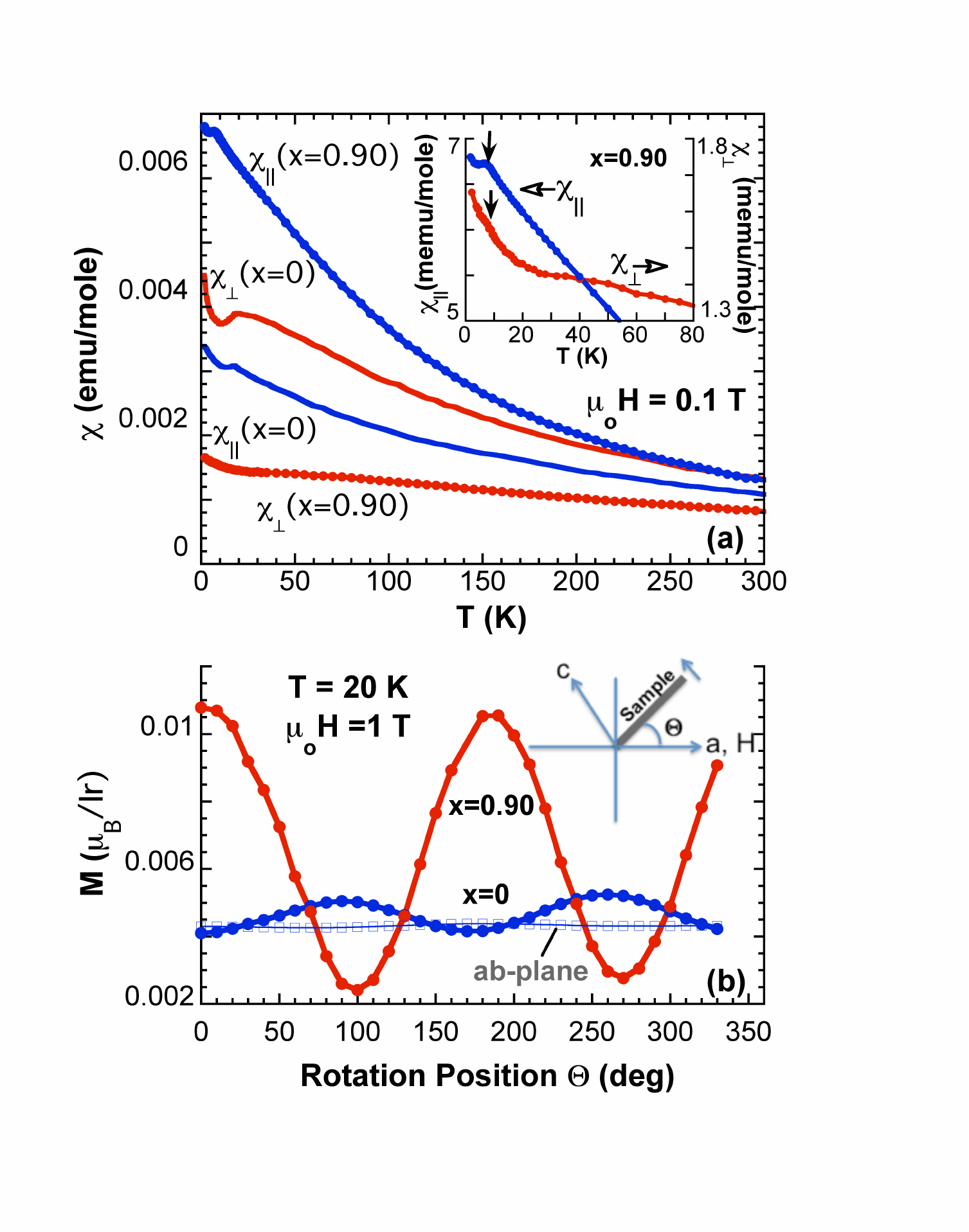}}
\caption{Comparisons between $x=0$ and $x=0.90$: The temperature dependence of (a) the in-plane and perpendicular-to-plane magnetic susceptibility, $\chi_\parallel(T)$ and $\chi_\perp(T)$ at $\mu_0 H = 0.1$ T for $x=0$ and $0.9$. {\em Inset:} The enlarged low-temperature $\chi_\parallel(T)$ (left scale) and $\chi_\perp(T)$ (right scale). (b) The angular dependence of magnetization M at $\mu_0 H = 1 T$ for $x=0$ and $0.90$ of \nlio~in the $a$-$c$-plane (solid circles) and for $x=0$ in the $a$-$b$-plane (empty squares); Inset: A schematic of the sample orientation relative to the $a$ and $c$ axis and the magnetic field $H$.  A detailed interpretation of the anisotropies of $\chi$ and its relation to the trigonal field, $\Delta$, is given in the text. Data for other dopings are shown in the SM. }
\label{fig:anis}
\end{figure}

{\em Anisotropy in $\chi_\parallel,\chi_\perp$: --} 
There are a number of striking features observed in our measurements of the direction dependent susceptibility in our single crystal samples, data for $x=0$ and $x=0.9$ are shown in Fig.~\ref{fig:anis}. Firstly, there are large anisotropies in the susceptibility even when $T\gg T_N$; indeed the Curie constant, ${\cal A}$ itself depends on the direction of the applied field. We define the Curie-constant by the usual definition, $\lim_{T\rightarrow \infty}\chi_{\parallel,\perp} = {\cal A}_{\parallel,\perp}/T$.  It is natural to attribute such anisotropies to the spin-orbit coupling (we study this in detail below). Secondly, the anisotropy between $\chi_\parallel(T)$ and $\chi_\perp(T)$ is reversed upon Li doping:  for $x=0$, ${\cal A}_\parallel < {\cal A}_\perp$ but for $x=0.9$, ${\cal A}_\parallel > {\cal A}_\perp$.

\begin{figure}[t]
\centerline{\includegraphics[trim = 0mm 25mm 5mm 10mm, clip,width=\columnwidth]{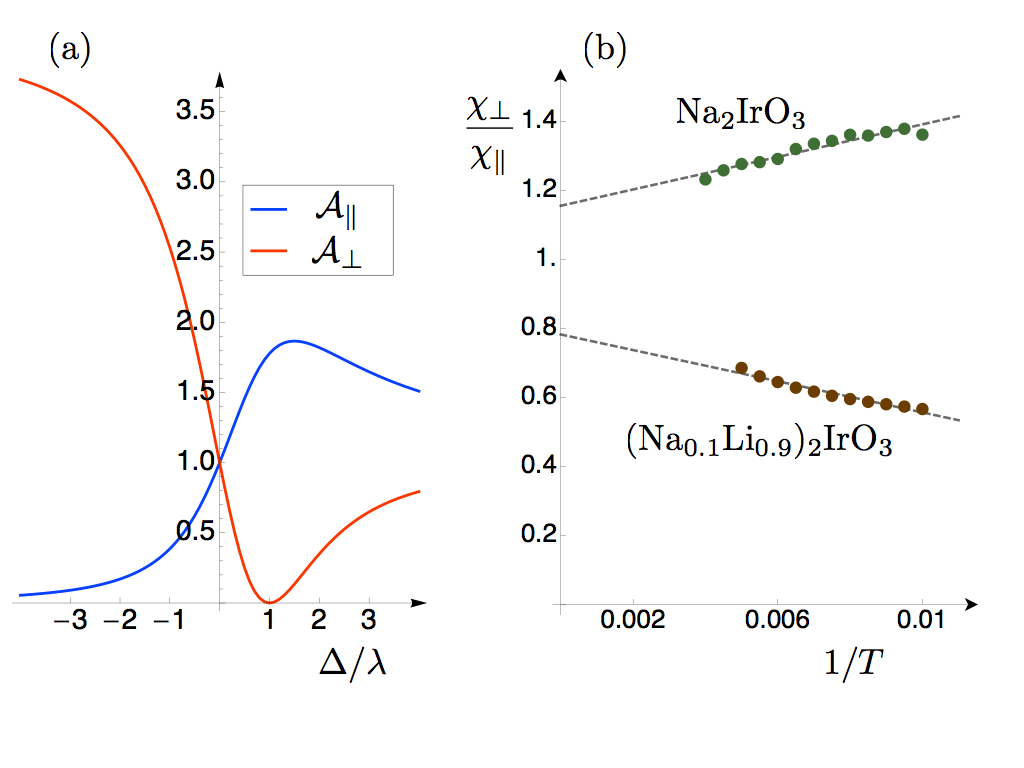}}
\caption{(color online). (a) Curie constants ${\cal A}_\parallel$ and ${\cal A}_\perp$
 as a function of the parameter $\Delta/\lambda$ calculated from the model Eq.~(\ref{eq:ham}). 
(b)  An extrapolation of the experimental
$\chi_\perp(T)/\chi_\parallel(T)$ in the high-temperature limit. At
$T=\infty$ this ratio should simply be  ${\cal A}_\perp/{\cal A}_\parallel$. Note that \nio~has ${\cal A}_\perp/{\cal A}_\parallel>1$ and hence $\Delta<0$, and based on the shown extrapolation for $x=0.9$, \lio~will extrapolate to ${\cal A}_\perp/{\cal A}_\parallel<1$ with $\Delta>0$. In both cases, clearly $\lambda \gg |\Delta|$. }
\label{fig:theory}
\end{figure}

In order to understand the origin of this change,  we calculate the $\chi_{\parallel,\perp}$ from the Hamiltonian for a single Ir ion with a $t^5_{2g}$ configuration with spin orbit coupling $\lambda>0$, a trigonal distortion $\Delta$ and an external field $\vec{h}$:
\begin{equation}
\label{eq:ham}
H = -\lambda \vec {l}\cdot \vec {s} - \Delta (\vec{l}.\hat n)^2 -
\vec{h}\cdot (2\vec s - \vec l)
\end{equation}
where $\vec{l}$ are the usual spin-1 matrices and $\vec{s}$ are the usual
spin-1/2 matrices, satisfying $[l_x,l_y]=il_z$ and $[s_x,s_y]=is_z$. We have made use here of the well-known, $l=1$ description of the $t_{2g}$ states~\cite{abragam1970:leq1}. The trigonal distortion vector $\hat n$ must point along a body diagonal of a cube that circumscribes the oxygen octahedra.  In the material (see inset of Fig.~\ref{fig:intro}(a)) the direction perpendicular to the honeycomb planes indeed points along a body diagonal for all the oxygen octahedra and is the natural direction to associate with $\hat n$ (we will verify this assumption from an analysis of the magnetic data below; structural data included in the SM also verifies this assertion). We calculate ${\cal A}_{\parallel,\perp}$, which in our theoretical calculation (see Supplementary Materials for details) only depends on $\Delta/\lambda$, these are plotted in Fig.~\ref{fig:theory}(a). We make the following observations from our model calculations: because of the rotational symmetry, ${\cal A}$ is the same for all directions perpendicular to $\hat n$; when $\Delta=0$ the response is rotationally invariant ({\em i.e.} ${\cal A}_\parallel = {\cal A}_\perp$) even when $\lambda \neq 0$; the anisotropy between ${\cal A}_\perp$ and ${\cal A}_\parallel$ is reversed as the sign of $\Delta$ changes; and, as expected for $\Delta/\lambda \rightarrow +\infty$ the system becomes rotationally invariant again.

At high temperatures ($T\gg \theta_{\rm CW}$), the Ir ions contribute to the susceptibility independently and one can hence use the high-$T$ experimental data to extract the Curie constants, ${\cal A}$.  First of all, we find that the experimentally measured ${\cal A}$ is the same within our errors of analysis for different directions in the honeycomb plane, but is clearly different for the direction perpendicular to the honeycomb layers (shown for $x=0$ by the rotation experiments in Fig.~\ref{fig:anis}(b)); this fact substantiates our claim that the $\hat n$ vector is along the direction perpendicular to the honeycomb layers. Next, as noted above the anisotropy in the susceptibility requires a finite $\Delta$, indicating that this paramater cannot be neglected in models of these materials. From Fig.~\ref{fig:theory}, it is clear that the change in anisotropy between $x=0$ and $x=0.9$ indicates that the sign of $\Delta$ changes between \nio~and \lio. 
Quantitatively, we extract the ratio
${\cal A}_\perp/{\cal A}_\parallel$ by extrapolating $\chi_\perp/\chi_\parallel$ as a
function of $1/T$, as shown in Fig.~\ref{fig:theory}(b). For Na$_2$IrO$_3$ we can do this reliably. We estimate the $\Delta/\lambda=
-0.05$ for  Na$_2$IrO$_3$, which we note is smaller than previous estimates~\cite{mazin2012:na_bs,gretarsson2013:hc_xray}. Based on our data, we conclude that for
Li$_2$IrO$_3$ the sign of $\Delta$ changes and its magnitude is
somewhat larger: our best estimate gives,  $\Delta/\lambda\approx 0.1$.

In summary we present the first evidence that \nio~and \lio~have distinct magnetic orders, by studying the evolution of structural, thermodynamic and magnetic properties of \nlio~with $x$ across a phase transition at $x\approx0.7$. Two possible tuning parameters for the phase transition that we have identified are the crystal field splitting, $\Delta$ and the anisotropy of the distortion of the honeycomb layers, both of which change sign from the Na to Li compounds. It is likely that a competition between the two is required to explain the magnetic ordering. Exploring these issues is an exciting direction for future theoretical research.

The authors are thankful to Natasha Perkins and Feng Ye for useful discussions. This work was supported by the National Science Foundation under grants DMR-0856234, EPS-0814194, DMR-1265162, and DMR-1056536 (RKK,MT). GC also acknowledges the hospitality of the China High Magnetic Field Laboratory of the Chinese Academy of Sciences.

\bibliography{/Users/rkk/Physics/PAPERS/BIB/iridates}

\section{SUPPLEMENTARY MATERIALS}

\subsection {Methods}

\begin{figure}[t]
\centerline{\includegraphics[trim = 0mm 130mm 0mm 0mm, clip,width=\columnwidth]{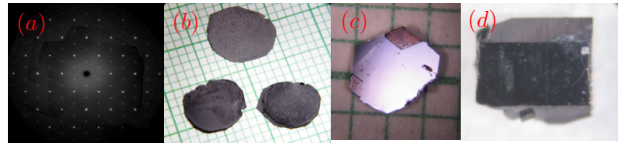}}
\caption{Single crystals of \nlio. (a) X-Ray diffraction pattern (hk0) for x = 0.9 (Li doping); (b,c,d) pictures of single crystals for x = 0, x=0.70 and x = 0.9 respectively.  }
\label{fig:xtal}
\end{figure}

Single crystals of \nlio~ $0\leq x \leq 0.90$ were grown using a self-flux method from off-stoichiometric quantities of IrO$_2$, Na$_2$CO$_3$ and Li$_2$CO$_3$. Similar technical details are described elsewhere~\cite{ye2012:nsc_na}. The pure Na$_2$IrO$_3$ crystals have a circular basal area corresponding to the honeycomb plane with diameters of more than 10 mm and thickness ~ 0.1 mm whereas \nlio~crystals are cylindrical-like with a hexagonal basal plane having diameters of ~ 2 mm and thickness ~ 2 mm (see Fig.~\ref{fig:xtal}). The structures of \nlio~were determined using a Nonius Kappa CCD X-Ray Diffractometer with sample temperature controlled using a nitrogen stream; they were refined by full-matrix least-squares method using the SHELX-97 programs~\cite{sheldrick2008:acta}. Chemical compositions of the single crystals were determined using both data of the  single-crystal X-Ray diffraction and energy dispersive X-ray analysis (EDX) (Hitachi/Oxford 3000). Electrical resistivity, magnetization and specific heat were measured using a Quantum Design MPMS7 SQUID Magnetometer and a Quantum Design Physical Property Measurement System with 14 T field capability.   

\subsection {Electrical Resistivity \& Other Measurements}

\begin{figure}[t]
\centerline{\includegraphics[trim = 10mm 25mm 5mm 30mm, clip,width=\columnwidth]{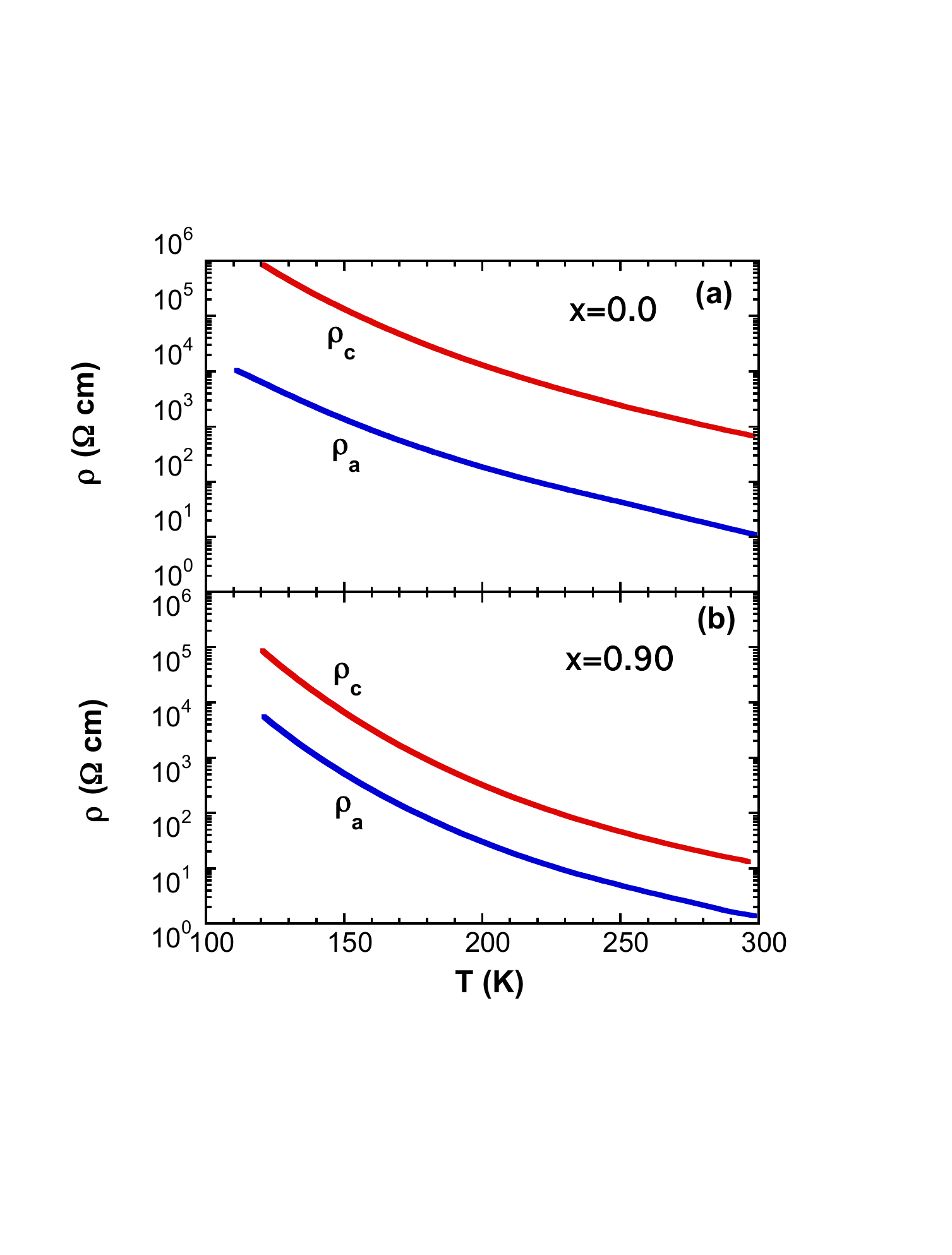}}
\caption{ The $a$-axis and $c$-axis electrical resistivity, $\rho_a(T)$ and $\rho_c(T)$ for (a) $x=0$, and (b) for $x=0.9$.  }
\label{fig:trans}

\end{figure}

\begin{figure}[t]
\centerline{\includegraphics[trim = 10mm 25mm 5mm 0mm, clip,width=\columnwidth]{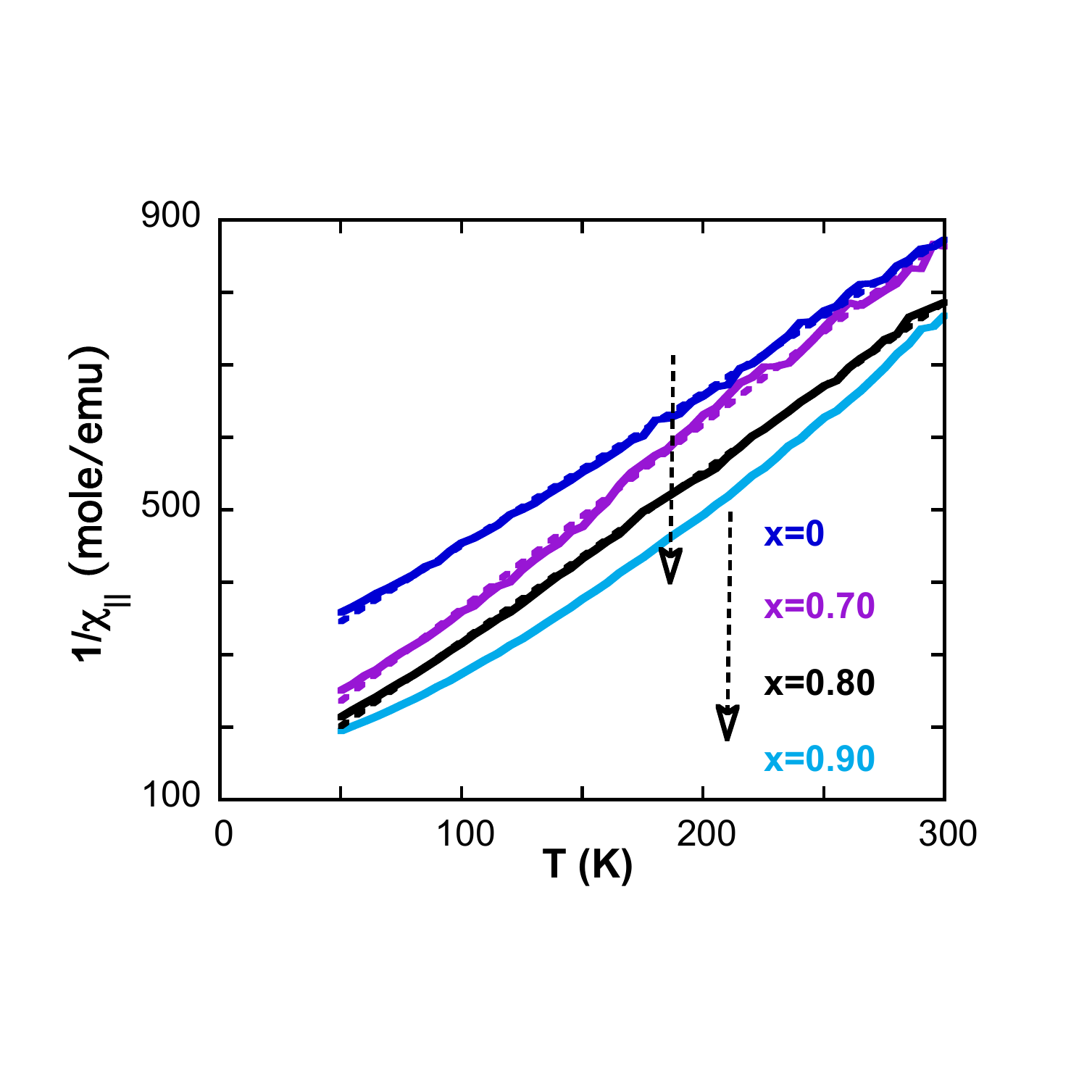}}
\caption{ Comparison of $1/\chi_\parallel$ vs $T$ for various different $x$ in \nlio.  }
\label{fig:suscinv}
\end{figure}

\begin{figure}[t]
\centerline{\includegraphics[trim = 10mm 40mm 5mm 0mm, clip,width=\columnwidth]{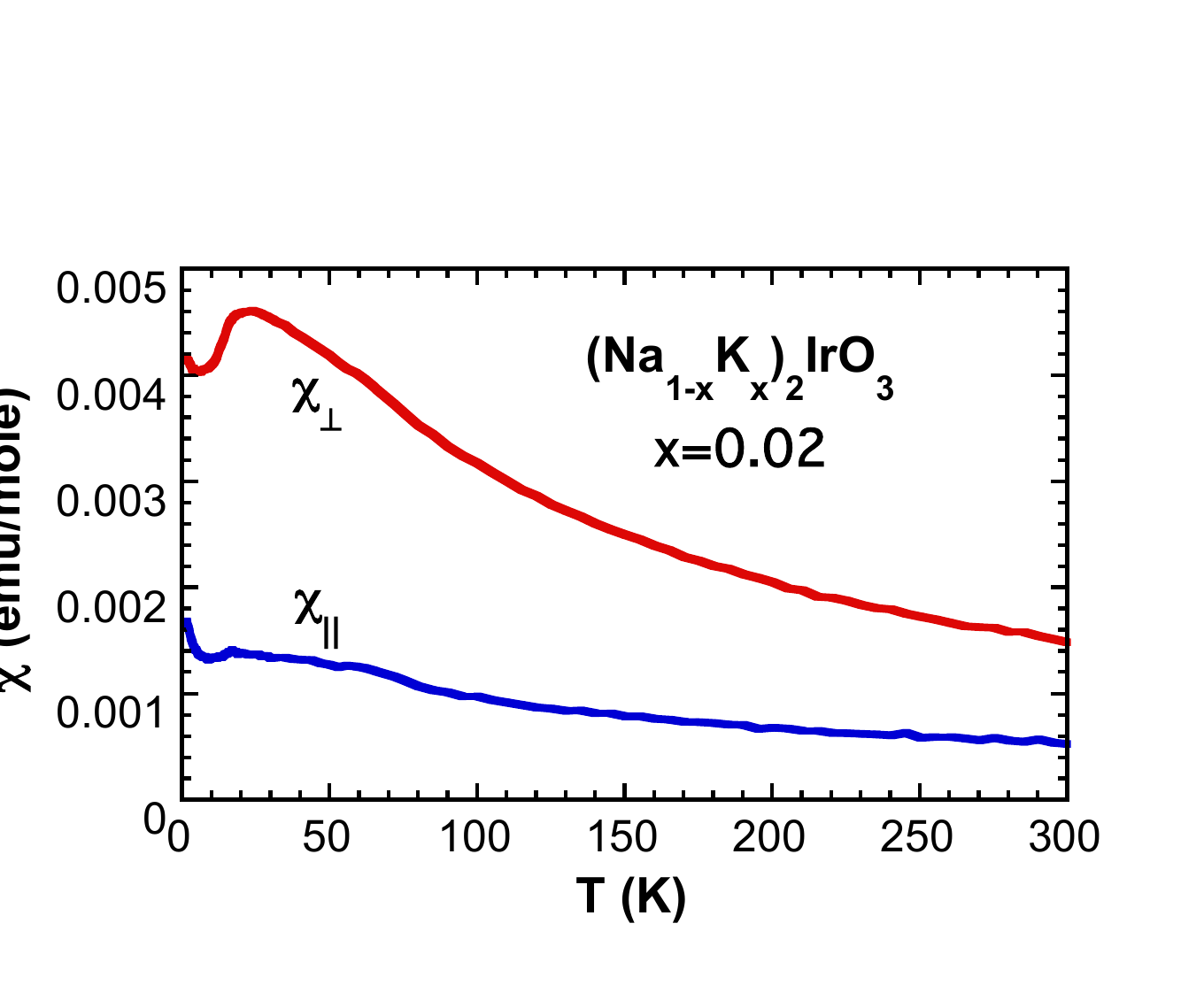}}
\caption{For completeness, we show $\chi_\parallel$ and $\chi_\perp$
  for K doped sample. Note that the anisotropy is same as in the pure
  Na sample and opposite from the Li doped samples. }
\label{fig:susckdope}
\end{figure}

{\em Transport:} Although not central to the results we have presented in the main text, we have also done simple transport studies on our samples. The anisotropy in the electrical resistivity is reduced, as shown in Fig.~\ref{fig:trans}.  $\rho_c$ is reduced by one order of magnitude whereas the $a$-axis resistivity $\rho_a$ remains essentially unchanged when $x$ increases from $x=0$ to $x=0.90$. The stronger $c$-axis compression clearly facilitates a greater $t_{2g}$-electron hopping between neighboring hexagons, and this explains the reduced $\rho_c$. In contrast, the nearly unchanged $\rho_a$ suggests that the highly anisotropic Ir-O-Ir hopping within hexagons stays the same despite the shortened $a$- and $b$-axis. 
 
{\em Specific Heat:} It deserves to be mentioned that for $x=0$, an additional anomaly in $C(T)$ is discerned at $T^* = 21$ K that is then followed by the zigzag magnetic order at $T_N = 18$ K [Fig.~2 (a)].  This additional anomaly could be an experimental manifestation of a two-phase transition predicated in a finite-temperature phase diagram of the classical Kitaev-Heisenberg model on the hexagonal lattice~\cite{price2012:kitH,price2013:long}. Between $T^*$ and $T_N$ there lies an intermediate phase with algebraically decaying correlations of the order parameter.  $T^*$ is not obviously discernable in $C(T)$ for Li doped \nio.

{\em Susceptibility:} We include a figure of $\chi_\parallel$ plotted
as $1/\chi_\parallel$ vs $T$ to show the evolution of the $\theta_{\rm
  CW}$, see Fig.~\ref{fig:suscinv}. We also show the $\chi_\parallel$
and $\chi_\perp$ for the K-doped sample, Fig.~\ref{fig:susckdope}.

\begin{figure}[!t]
\centerline{\includegraphics[trim = 50mm 10mm 55mm 0mm, clip,width=\columnwidth]{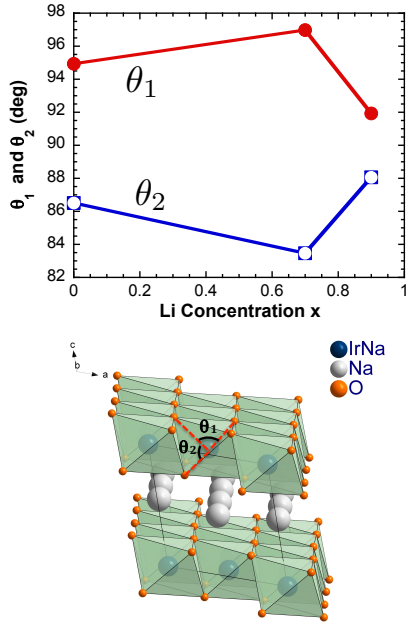}}
\caption{Structural data from X-ray refinement clearly shows a trigonal
  distortion for all $x$ in the honeycomb Iridates in a direction
  perpendicular to the planes. Both angles should be 90$^\circ$
for a perfect octahedron. The trigonal squeeze (because $\theta_1>\theta_2$) can be thought of as a
flattening of the octahedron perpendicular to the $a$-$b$ layer.  }
\label{fig:angles}
\end{figure}

\begin{figure}[t]
\centerline{\includegraphics[trim = 45mm 75mm 25mm 10mm, clip,width=\columnwidth]{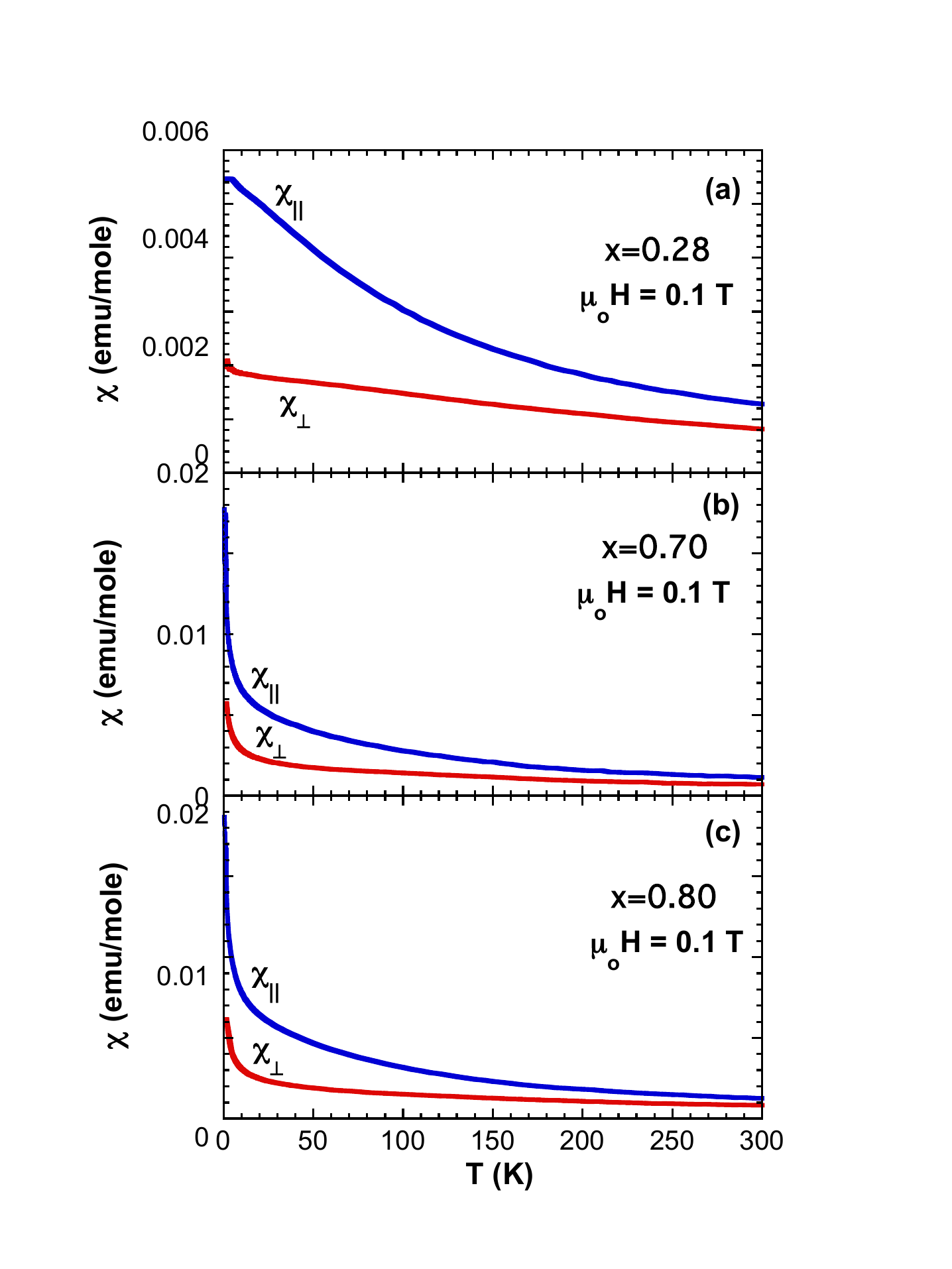}}
\caption{ $\chi_\perp$ and $\chi_\parallel$ as a function of $T$ for
  $x=0.28,0.70$ and 0.80. The data for $x=0,0.90$ is shown in the main
manuscript. Analogous to Fig. 5 of the manuscript, extrapolated values at high-$T$ of ${\cal A}_\perp/{\cal A}_\parallel$
are 0.63, 0.67, 0.71  (for $x=0.28,0.70$ and $0.80$)}
\label{fig:susc_pp}

\end{figure}

\subsection{Calculation of Curie Constants}

Here we provide a quick sketch of the calculation of the Curie constants, ${\cal A}_{\perp,\parallel}$ in Fig.~5 (a). Starting with Eq.~(1), we can diagonlize the Hamiltonian with ${\vec h}=0$. Since there are five electrons in total, the topmost Kramers degenerate pair will be singly occupied,
\begin{equation}
| \pm \rangle = \cos \left (\frac{\theta}{2}\right ) \left | \frac{3}{2},\pm\frac{1}{2} \right \rangle \pm \sin \left (\frac{\theta}{2}\right ) \left | \frac{1}{2},\pm\frac{1}{2} \right \rangle
\end{equation}
where $\cos \theta= \frac{-\frac{3\lambda}{4}+\frac{\Delta}{6}}{\sqrt{\left (\frac{3\lambda}{4}-\frac{\Delta}{6}\right )^2 +\frac{2 \Delta^2}{9}}}, \sin \theta= \frac{\frac{\sqrt{2}\Delta}{3}}{\sqrt{\left (\frac{3\lambda}{4}-\frac{\Delta}{6}\right )^2 +\frac{2 \Delta^2}{9}}}$ and the kets on the right are the $\jeff$ eigenstates of the form $\left | J, M_J \right \rangle$. Calculating the matrix elements of the magnetization operator, $\vec{l}-2 \vec{s}$ in these states, then gives us the effective $g$-factors from which we find,
\begin{eqnarray}
{\cal A}_\perp  = \frac{1}{4} (1 - \cos \theta - 2\sqrt{2}\sin\theta)^2\nonumber \\
{\cal A}_\parallel  = \frac{1}{4} (1 - \cos \theta + \sqrt{2}\sin\theta)^2
\end{eqnarray} 
where ${\cal A}_\perp$ is the Curie constant along the $\hat n$ direction and ${\cal A}_\parallel$ is the susceptibility perpendicular to the $\hat n$ direction.
We note here that the anisotropy of the Curie constants can equivalently be interpreted as arising from the $g$-factor becoming direction dependent ({\em i.e.}, becoming a $g$-tensor).

\end{document}